# Pulsar Constellation Evolution and PPP Convergence using LEO GNSS

Claire Mah, Thyagaraja Marathe, Ananya Vishwanath, Matteo Gala, Tyler G. R. Reid, Qamar Bader
*Xona Space Systems*

## BIOGRAPHIES

**Claire Mah** is a GNSS Solutions Specialist at Xona. She received her master's degree from the University of Calgary in geomatics engineering for research focusing on GNSS positioning and estimation for low earth orbit satellite systems.

**Dr. Thyagaraja Marathe** is a Principal Engineer at Xona, specializing in PNT using LEO satellite constellations. He previously worked at Rx Networks and Accord Software on GNSS corrections and receiver development. He holds a Ph.D. in Geomatics Engineering from the University of Calgary, where he also completed postdoctoral research. His interests include LEO PNT, high-accuracy positioning, and GNSS receiver design. He also holds degrees from BITS Pilani and VTU, India.

**Ananya Vishwanath** is a UE Verification Specialist at Xona. She received her MSc in Geomatics Engineering from the University of Calgary, focusing on precise GNSS positioning using low-cost receivers and smartphones.

**Matteo Gala** is a GNSS Engineer at Xona where he focuses on evaluating the positioning and timing performance of Xona's Pulsar low-Earth-orbit satellite navigation system. He has previously worked on GNSS and PNT research and development in both academic and industry settings, with experience spanning navigation system analysis and performance assessment. Matteo received his MSc in Mechatronics from Cranfield University and BEng in Aerospace Engineering from The University of Nottingham.

**Dr. Tyler G. R. Reid** is co-founder and CTO of Xona. Previously, Tyler worked as a Research Engineer at the Ford Motor Company in localization and mapping for self-driving cars. He has also worked as a software engineer at Google and as a lecturer at Stanford University, where he co-taught the GPS course. Dr. Reid received his PhD (2017) and MSc (2012) in Aeronautics and Astronautics from Stanford, where he worked in the GPS Lab.

**Dr. Qamar Bader** is a GNSS Specialist at Xona. She received her Ph.D. (2026) and M.A.Sc. (2023) degrees in electrical and computer engineering from Queen's University, Canada. Prior to this, she obtained her B.Sc. degree in electrical engineering from Qatar University (2016). Her research interests include LEO PNT, sensor fusion, and state estimation.

## ABSTRACT

The objective of this paper is to demonstrate the future impact of a dedicated low-earth-orbit (LEO) Global Navigation Satellite System (GNSS) on positioning, navigation, and timing (PNT), in addition to a reflective discussion on how common GNSS evaluation methods can best reflect LEO satellite dynamics. This paper will feature Xona's LEO constellation, Pulsar, at various stages of deployment in a case study comparing GPS, LEO-augmented GPS, and standalone LEO. Results include visibility and dilution of precision (DOP) analysis for Pulsar at four stages of constellation deployment, as well as two cases exploring how precise point positioning (PPP) convergence is affected by Pulsar's LEO satellite dynamics.

Pulsar is highlighted in this paper for its purposeful design that allows for both GNSS augmentation through compatible signals and GNSS independence (Gala et al., 2026). As dedicated LEO satellite-based positioning becomes more prominent in the PNT industry, it is important to consider which methods are used to evaluate constellation geometry and positioning performance. By evaluating the constellation through a variety of different methods in this paper, a discussion on the future evolution of LEO navigation constellations is offered. Pulsar is compared to MEO GNSS, both as an augmentation and a standalone service, through availability, dilution of precision, and PPP convergence.

## 1. INTRODUCTION

Following the success of Xona's in-orbit-validation (IOV) satellite, Pulsar-0, this paper will take a detailed look at the future of the Pulsar constellation. Over the next few years, Xona is planning to roll out its constellation in several stages. The first operational stage will include 16 satellites in inclined orbital planes that will consistently provide 1 satellite in view at most mid-latitude locations across North America, Europe, Japan, and Korea. This marks the minimum viable product (MVP) for Pulsar. The next major stage is at 48 satellites deployed in inclined orbits, giving rise to high availability of several satellites across a broader geography. At 192 satellites in inclined orbits, this offers 4 or more satellites in view globally for positioning, except for at high latitudes and the poles. Finally, the complete constellation of 258 satellites will provide global coverage with an additional polar orbiting layer, ensuring visibility of at least 4 satellites at all locations.

In a signal compatibility study, Reid et al. (2025) demonstrated that the fully deployed Pulsar constellation can coexist with existing GNSS constellations in L-band without degrading already deployed GNSS signals. In addition to simply coexisting, Pulsar's carrier frequencies allow for backwards compatibility with most GNSS receiver hardware (Reid et al., 2025) (Leclère et al., 2025). With compatibility established, this study evaluates Pulsar based on the performance of GNSS with Pulsar augmentation at several stages of constellation maturity, as well as standalone Pulsar performance. Preliminary results from Xona's IOV satellite, Pulsar-0. will also be featured. Pulsar is expected to complement GNSS with its adjacent frequency, strong signals, and dynamic geometry, as demonstrated using PPP convergence simulations.

PPP uses undifferenced carrier phase observations to estimate a precise position, enabled by correction services but requiring long convergence times (Kouba et al., 2017). The convergence time depends on atmospheric effects, multipath conditions, satellite geometry, and algorithm design. With the introduction of Pulsar and other LEO-PNT constellations, PPP convergence and carrier phase ambiguity resolution will be accelerated due to the fast-changing geometry of low orbit satellites (Zheng et al., 2024). Early on-orbit measurements from Pulsar-0 exhibited pseudorange residuals with an RMS of 43 mm over the reported test interval (Gala et al., 2026), demonstrating low short-term code-domain measurement scatter. However, the published analysis does not independently quantify satellite orbit and clock errors relative to defined terrestrial and time references. Consequently, the reported value should not be interpreted as a formally established SISRE or system-level URE. Although the result indicates promising ranging precision, a like-for-like comparison with GPS would require equivalent measurement definitions, reference products and statistical treatment. In addition to centimeter-level ephemeris for Pulsar satellites, Xona will also offer MEO GNSS corrections including orbit and clock corrections as well as code and phase biases via broadcast on X5, all supporting faster PPP convergence. Broadcast ephemeris will be distributed with a data rate of over 400 bps on X1 and over 3000 bps on X5, compared to 50 bps on L1C/A and 100 bps on L5, reducing time to first fix (TTFF). All factors taken together, Pulsar has the potential to improve multiple aspects of PPP performance. This paper will focus on PPP convergence from the perspective of the additional availability and dynamic satellite geometry provided by a LEO constellation.

## 2. METHODOLOGY

This section details the workflow and the evaluation metrics that are used to discuss Pulsar's availability, geometric strength, and PPP convergence.

### 2.1 Global Coverage and Availability

Global satellite availability maps have been developed based on the orbital parameters of Pulsar (Leclère et al., 2025), including earlier stages of constellation deployment at 16, 48, 192, and 258 satellites deployed.

At full operational capability (FOC), Pulsar will consist of 258 satellites, including 12 planes of 16 satellites with 53° inclination and 6 planes of 11 satellites with 97.5° inclination. The minimum viable product (MVP) for Pulsar is defined as 16 satellites, consisting of 2 inclined planes with 8 satellites each. All orbit configurations listed in Table 1 have a nominal altitude of 1080 km and represent circular orbits. The satellite planes are offset in their mean anomaly phasing to optimize ground coverage. In this analysis, satellite spacing and orbital plane spacing are assumed to be evenly distributed throughout earlier stages of deployment, although the actual spacing may depend on satellite launch availability and other factors.

**TABLE 1**

Nominal Pulsar Constellation Characteristics (Leclère et al., 2025)

| | Minimum Viable Product | Partially Populated Inclined Planes | | Fully Populated Inclined Planes | Full Operational Capability | |
|---|---|---|---|---|---|---|
| **Number of orbital planes** | 2 | 4 | | 12 | 12 inclined | 6 polar |
| **Orbital planes RAAN offset** | 180° | 90° | | 30° | 30° | 60° |
| **Number of satellites per plane** | 8 | 2 × 8 | 2 × 16 | 16 | 16 | 11 |
| **Satellites mean anomaly offset** | 45° (360° / 8) | 45° (360° / 8) | 22.5° (360° / 16) | 22.5° (360° / 16) | 22.5° (360° / 16) | ≈ 32.7° (360° / 11) |
| **Orbit inclination** | 53° | 53° | | 53° | 53° | 97.5° |

These parameters are used in simulations to determine the global availability of Pulsar. Simulations were run over 24-hour periods and show how often a user in each location would expect to have at least 1, 2, 3, or 4 Pulsar satellites visible above a minimum elevation angle of 5°, 10°, or 15°. Note that in this paper, the term availability is used to refer to line-of-sight visibility to a satellite, as opposed to service availability. For example, 95% geometric availability does not equate to 95% service uptime after accounting for variables such as space system reliability, ground system reliability, or environmental factors.

The simulations also analyze the dilution of precision (DOP) of the satellite geometry at each location. The 3D position DOP (PDOP) and horizontal DOP (HDOP) are calculated at each epoch according to (Alghisi et al., 2025):

$$\boldsymbol{H}_{(n\times u)} = \begin{bmatrix} \frac{\partial h_1(\hat{x})}{\partial x_1} & \cdots & \frac{\partial h_1(\hat{x})}{\partial x_u} \\ \vdots & \ddots & \vdots \\ \frac{\partial h_n(\hat{x})}{\partial x_1} & \cdots & \frac{\partial h_n(\hat{x})}{\partial x_u} \end{bmatrix} \tag{1}$$

$$\boldsymbol{N} = \boldsymbol{R}_{\boldsymbol{ENU}}^{\boldsymbol{ECEF}} [\boldsymbol{H}^T \boldsymbol{H}]_{(3\times3)} \boldsymbol{R}_{\boldsymbol{ENU}}^{\boldsymbol{ECEF}^T} \tag{2}$$

$$PDOP = \sqrt{\boldsymbol{N}_{0,0} + \boldsymbol{N}_{1,1} + \boldsymbol{N}_{2,2}} \tag{3}$$

$$HDOP = \sqrt{\boldsymbol{N}_{0,0} + \boldsymbol{N}_{1,1}} \tag{4}$$

where $H$ is the Jacobian matrix of the observation equations, $n$ is the number of observations, $u$ is the number of parameters to be estimated, $N$ is the normal matrix, and $\boldsymbol{R}_{\boldsymbol{ECEF}}^{\boldsymbol{ENU}}$ is the rotation matrix from the ECEF frame to a local ENU frame. We assume 4 observations from each visible satellite, including one pseudorange and one carrier phase from each signal frequency L1C/A and L5 on GPS, or X1 and X5 on Pulsar.

Global maps of satellite availability are indexed using H3 cells (Sahr, 2011) and are used to effectively visualize trends in satellite visibility and constellation coverage. H3 indexing is a standardized geospatial grid system that scales from 4 million $km^2$ cells to less than 1 $m^2$ cells, depending on the application. We use resolution level 2, where the globe is divided into 5882 total cells, each covering approximately 87 000 $km^2$. User locations are simulated at the center coordinates of each cell, assuming an altitude of zero meters.

### 2.2 PPP Convergence Simulations

PPP convergence is evaluated using two different methods. In Part 1, Pulsar is simulated entirely in software, and generated observations are processed through an Extended Kalman filter (EKF) to evaluate the position error for a global set of users at each stage of constellation deployment. In Part 2, real data collected from Pulsar-0 is processed through a modified RTKLIB (Takasu & Yasuda, 2009) implementation and the covariance convergence is compared between GPS-only and GPS augmented by Pulsar-0. In both cases, Pulsar is modelled to have the same precision as GPS under the assumption that the observations have already been corrected, allowing for a direct comparison of the constellations, focused on satellite dynamics and geometry.

#### 2.2.1 Part 1: PPP Convergence from Global Simulations

To predict the behaviour of a GNSS positioning engine with Pulsar, simulation software was written to perform orbit propagation, generate measurements, and process them through an EKF. The dynamic movement of LEO satellites significantly improves the geometric network strength of the positioning solution (Murata et al., 2024). Coupled with more precise observations from Pulsar, PPP convergence is expected to be faster and more precise when compared to MEO GNSS. The EKF model that was used in this simulation is summarized by the following set of equations:

$$\hat{\boldsymbol{x}} = \begin{bmatrix} x \\ y \\ z \\ cdT \\ ZTD \\ ISB \\ G_N \\ G_E \\ N_{S1,f1} \\ N_{S1,f2} \\ \vdots \\ N_{Sn,f1} \\ N_{Sn,f2} \end{bmatrix} \tag{5}$$

$$\boldsymbol{K} = \boldsymbol{P}_k^- \boldsymbol{H}_k^T (\boldsymbol{H}_k \boldsymbol{P}_k^- \boldsymbol{H}_k^T + \boldsymbol{R}_k)^{-1} \tag{6}$$

$$\boldsymbol{P}_k = (\boldsymbol{I} - \boldsymbol{K}_k \boldsymbol{H}_k) \boldsymbol{P}_k^- \tag{7}$$

where $\boldsymbol{K}$ is the Kalman gain matrix, $\boldsymbol{P}_k$ is the covariance matrix of the state vector $\hat{\boldsymbol{x}}$, and $\boldsymbol{R}_k$ is the noise covariance matrix of the observations. $\hat{\boldsymbol{x}}$ is the state vector containing all the parameters to be estimated including 3D position coordinates, receiver clock bias, troposphere zenith delay, inter-system bias (ISB) for each constellation less the reference one, north and east gradients, and carrier phase ambiguities for each frequency for satellites 1 through *n*.

For the global simulations, observations were generated, with a 5° elevation mask and a 1 Hz data rate, for a set of 176 static user locations distributed across all latitudes. With these user observations, the PPP filter was run once per hour for 24 hours for a duration of 400 s, resulting in over 4000 simulated PPP scenarios for each combination of constellations. In this analysis, GPS + Galileo is used as the baseline for comparing MEO GNSS to Pulsar augmented systems. Of particular interest are the comparisons of standalone GPS vs standalone Pulsar, and GPS + Galileo vs GPS + Galileo + Pulsar. All simulations used float ambiguities and dual frequencies; simulations involving GPS used L1C/A and L2, Galileo used E1 and E5, and Pulsar used X1 and X5.

#### 2.2.2 Part 2: PPP Convergence from Pulsar-0 with RTKLIB

For a second look at PPP convergence, live sky observations from Pulsar-0 were collected and processed through RTKLIB, which was modified to accept observations and ephemeris from Pulsar. RTKLIB is an open-source GNSS processing software (Takasu & Yasuda, 2009) that has a known and accepted positioning engine that can be used to process Pulsar. However, note that the results and analysis contained in this section should be considered preliminary, and work on positioning with live Pulsar data and this RTKLIB implementation is still in progress.

Starting from the Demo5 version of RTKLIB (*RTKLIB - Demo5 User Manual*, 2024), the program was modified to recognize and parse Pulsar in RINEX observation and navigation files and added capability to process Pulsar measurements in the same manner as all other GNSS constellations. The RINEX format also required a slight modification to accommodate 3-digit satellite vehicle identifiers (SVID) for Pulsar by utilizing hexadecimal notation. No modifications were made to the positioning algorithms.

A recorded signal from an IOV pass over Xona's Montreal office on August 23rd, 2026, was processed using an in-house software defined receiver that demodulated and decoded ephemeris messages including satellite positions for GPS and Pulsar-0. RINEX navigation and observation files were generated that could then be used in post-processed PPP algorithms and analysis, like RTKLIB.

Pulsar-0 has different orbital parameters than the full constellation. It orbits at 520 km with a polar inclination of 97° (Leclère et al., 2025). This IOV satellite pass peaked at 68° elevation and was recorded to have a carrier to noise density ratio of 57 dB-Hz. The data was processed for GPS only, and for GPS + Pulsar-0 using the same settings and parameters, as given in Table 2 below. The process noise is set to RTKLIB's default value, but the carrier phase error is increased to 5 cm to allow the filter more leniency in accepting observations. Observations are weighted proportionately to the satellite elevation angle, and pseudorange errors are 100 times larger than carrier phase errors. Note that these settings come from very early testing and any positioning engine integrating LEO observations will likely require further filter tuning.

**TABLE 2**

RTKLIB Settings

| Parameter | Value |
|---|---|
| **Positioning Mode** | PPP-static |
| **Frequencies** | L1C/A (for GPS), X1 (Pulsar) |
| **Elevation Mask** | 5° |
| **Ambiguity Resolution Mode** | Float |
| **Carrier Phase Error** | 0.05 m |
| **Elevation-Dependent Carrier Phase Error** | 0.003 m |
| **Pseudorange Error Ratio** | 100 |
| **Process Noise Bias** | 0.0001 m |

Due to challenges in properly modelling and estimating satellite clock biases and other errors present in live data from a single satellite source, this study will use a covariance analysis as a preliminary method to evaluate PPP convergence with live data. From the EKF equations above, the covariance matrix, $\boldsymbol{P}_k$, is estimated based only on satellite positions and measurement variances, independent of the state estimates or any observations. Precision metrics are derived from the diagonal elements of $\boldsymbol{P}_k$ (Equation 7) corresponding to the position states, rotated into a local east, north, up (ENU) frame as:

$$\sigma_{3D} = \sqrt{\sigma_E^2 + \sigma_N^2 + \sigma_U^2} \tag{8}$$

$$\sigma_{horizontal} = \sqrt{\sigma_E^2 + \sigma_N^2} \tag{9}$$

This is analogous to the DOP (equations (3) and (4)), propagated over time to reflect the measurement precision and the convergence of the filter towards a more precise solution. While DOP is a good indicator of satellite geometry and visibility, its instantaneous nature neglects the added benefits of the dynamic geometry of LEO satellites in the EKF (Mah & O'Keefe, 2025). Therefore, horizontal and 3D variance are useful metrics in evaluating the impact of LEO satellite geometry.

## 3. RESULTS AND DISCUSSION

Figure 1 shows Pulsar at various stages of deployment, from the minimum viable product at 16 satellites to full operational capability at 258 satellites. This design uses a combination of inclined and polar planes to ensure global coverage at all latitudes with earlier stages concentrating satellite visibility around mid-latitudes and major population centers.

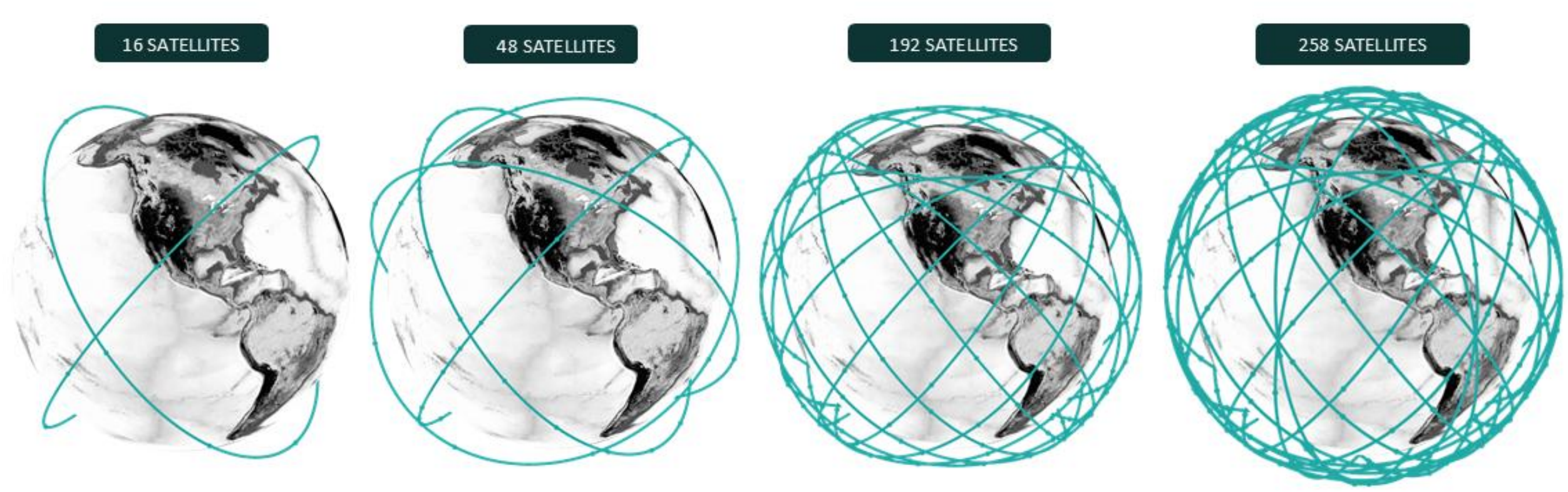

**FIGURE 1** Pulsar Constellation Deployment Stages

## 3.1 Global Coverage and Availability

All satellite availability maps in this section use the same colour scale, skewed for higher resolution at higher levels of availability. In general, orange cells indicate intermittent coverage, yellow and green cells correspond to persistent coverage, and blue cells denote consistent and reliable coverage. The inclined planes are designed to provide coverage at low and mid latitudes, including most populated landmasses. The polar planes then fill in the gaps at the north and south poles, as well as provide some redundancy to the inclined plane layer.

### *3.1.1 Minimum Viable Product: 16 Pulsar Satellites*

Figure 2 shows the availability of 1 visible Pulsar satellite above a 5° elevation mask with only 16 satellites on orbit. This stage offers intermittent visibility in most populated areas. These satellites orbit with 53° inclination and therefore provide the best coverage in the mid-latitude bands. With this coverage, Pulsar will be suitable for applications including precise time transfer, intermittent positioning including stationary users indoors (Marathe et al., 2026), and GNSS positioning augmentation.

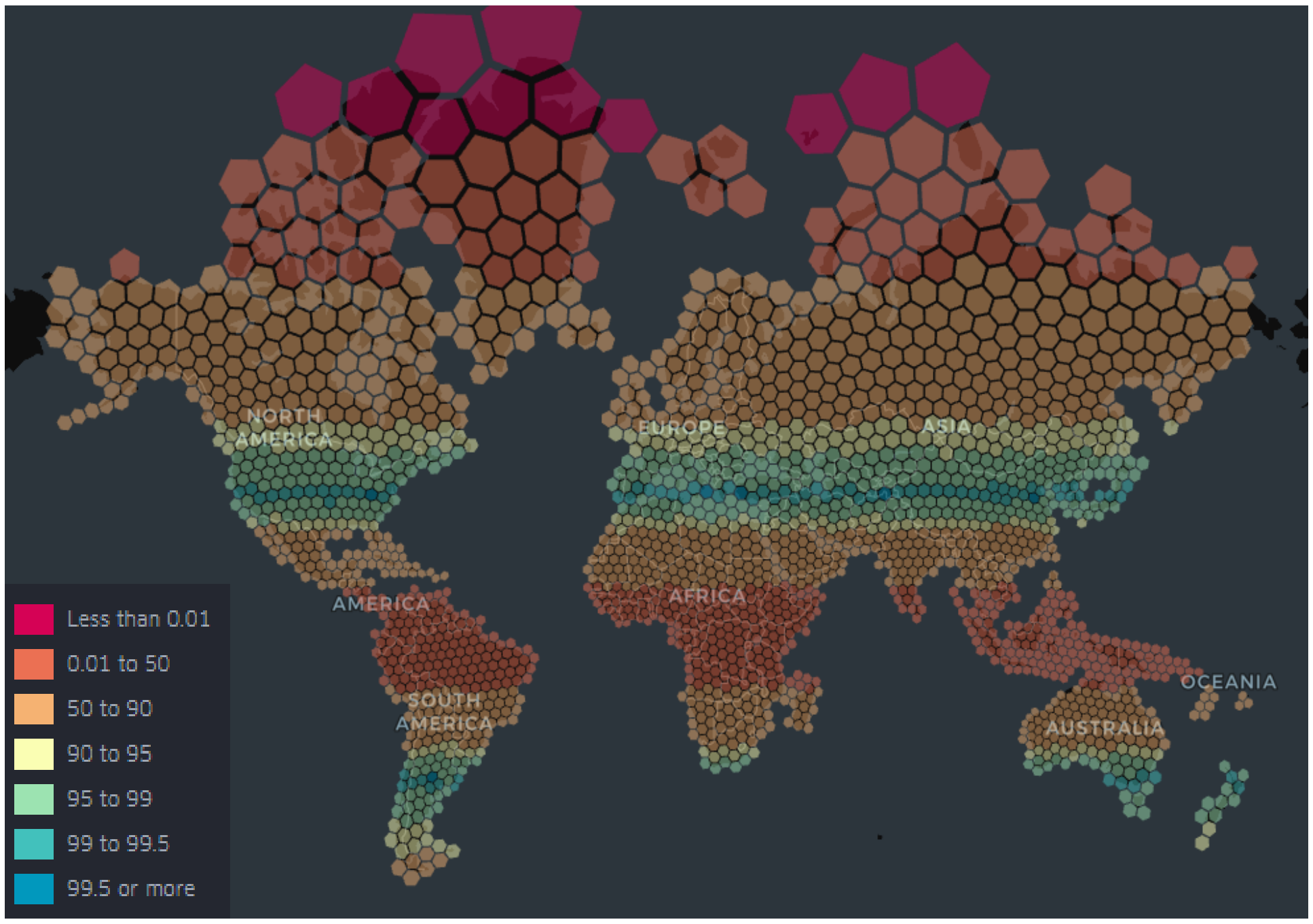

**FIGURE 2** Availability (%) of 1 Visible Pulsar Satellite Above 5° Elevation - 16 Satellites

On average, LEO satellites are observed at lower elevation angles compared to MEO GNSS constellations. A detailed explanation and analysis of the satellite distribution by elevation is provided by Leclère et al. (2025). Figure 2 and Figure 3 show the same constellation simulation with 5° and 10° masks, respectively. With a 5° elevation mask, many locations in the mid-latitude regions will have at least 1 satellite visible 95% of the time or more (green and blue cells). However, once the elevation mask is raised to 10°, this availability does not exceed 90% globally. Thus, these figures highlight the challenge that LEO constellations face in providing sufficient coverage, particularly in urban environments where buildings often mask 30° or higher.

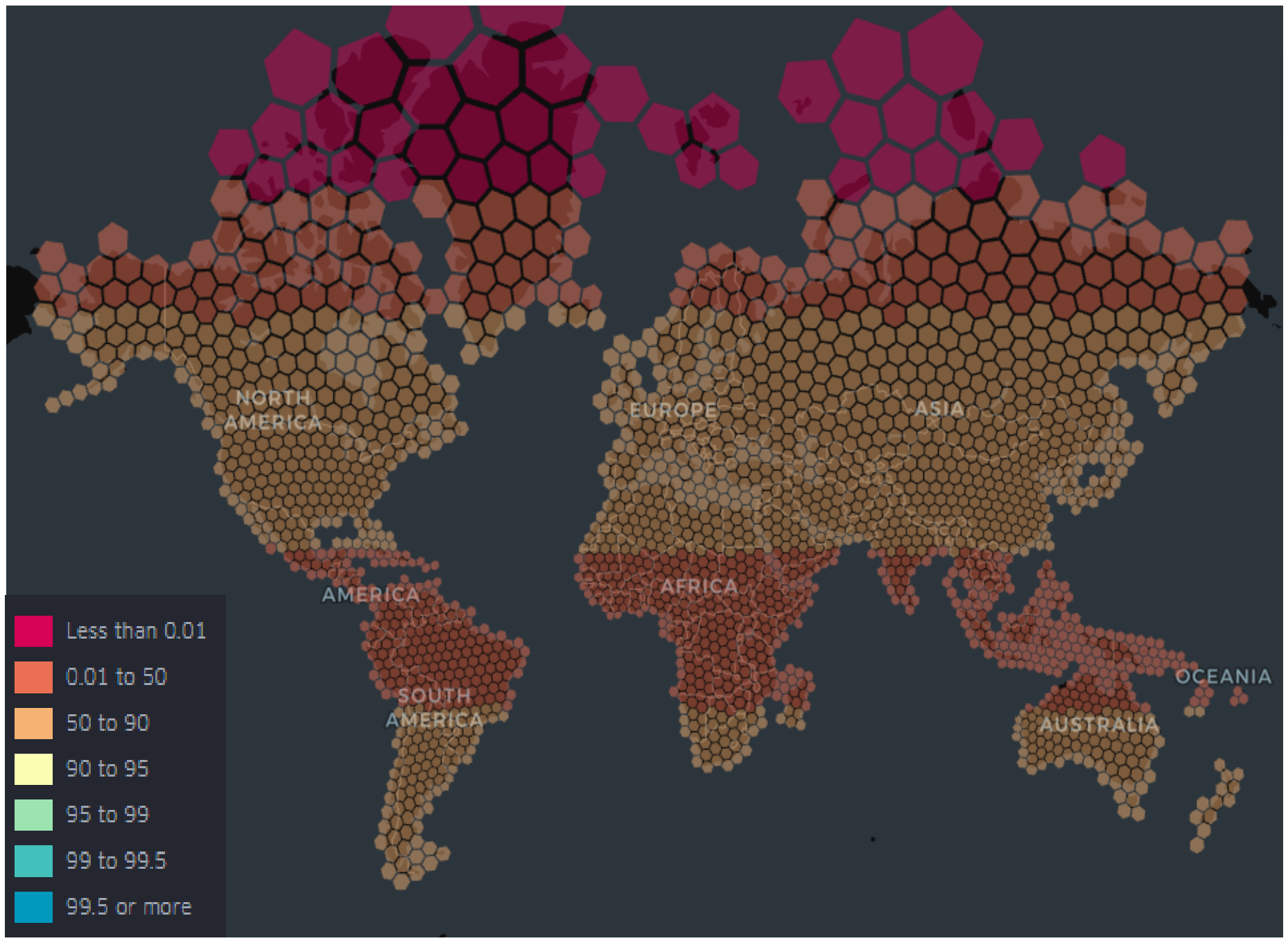

**FIGURE 3** Availability (%) of 1 Visible Pulsar Satellite Above 10° Elevation - 16 Satellites

### *3.1.2 Partially Populated Inclined Planes: 48 Pulsar Satellites*

At 48 satellites, Pulsar's coverage is significantly improved, with latitudes between approximately ± (20° to 65°) having at one or more satellites in view at least 99.5% of the time. Due to the inclination of the orbital planes, the best coverage and the most overlap occurs in the mid-latitude regions. Regions in the far north, far south, or near the equator will experience longer gaps in coverage.

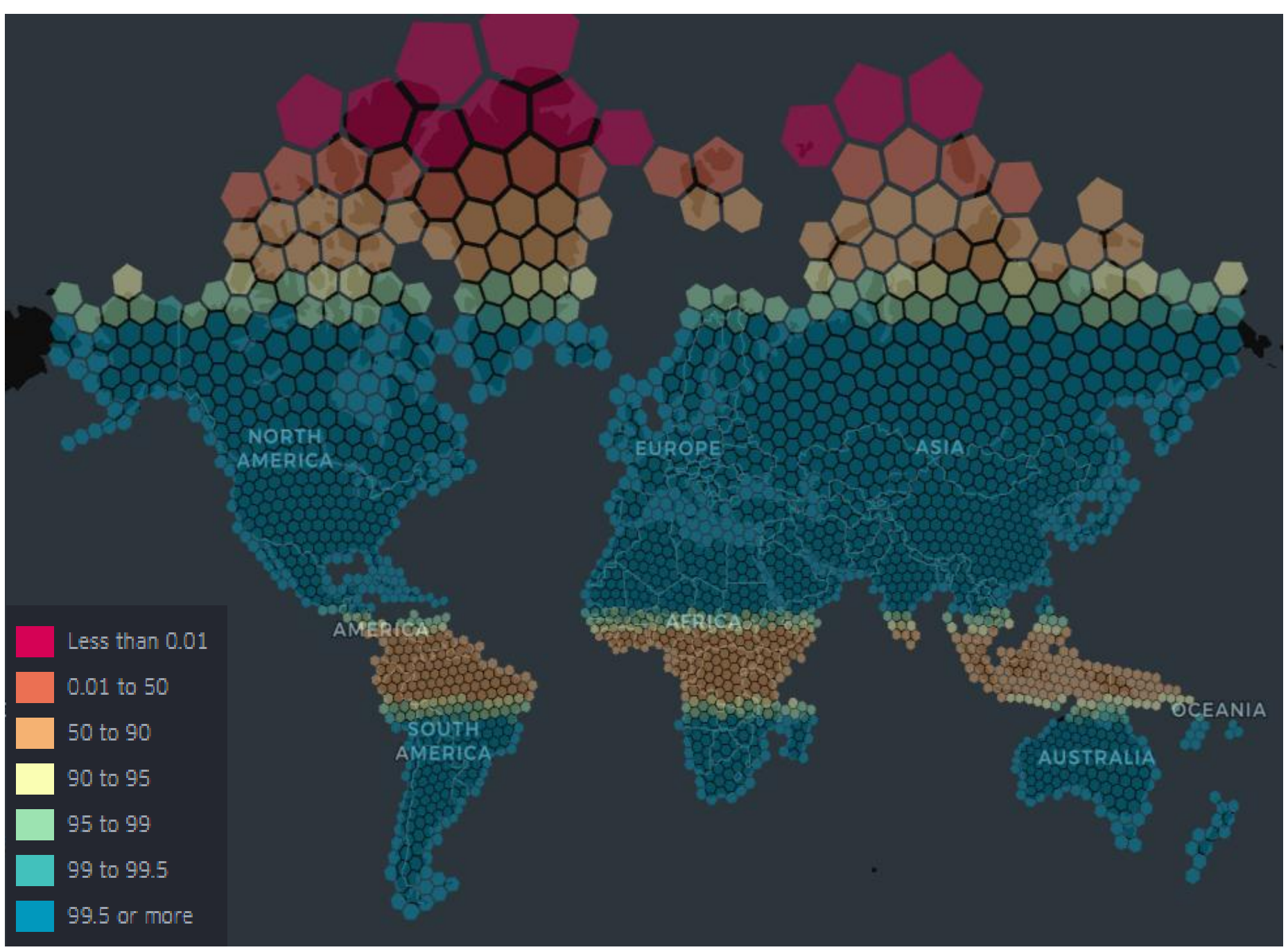

**FIGURE 4** Availability (%) of 1 Visible Pulsar Satellite Above 5° Elevation - 48 Satellites

### *3.1.3 Fully Populated Inclined Planes: 192 Pulsar Satellites*

At 192 satellites, the 12 inclined planes are fully populated. Figure 5 shows that at least 4 Pulsar satellites will be visible above 5° elevation 99.5% of the time, at all latitudes from -70° to +70°, thus covering all major populated locations.

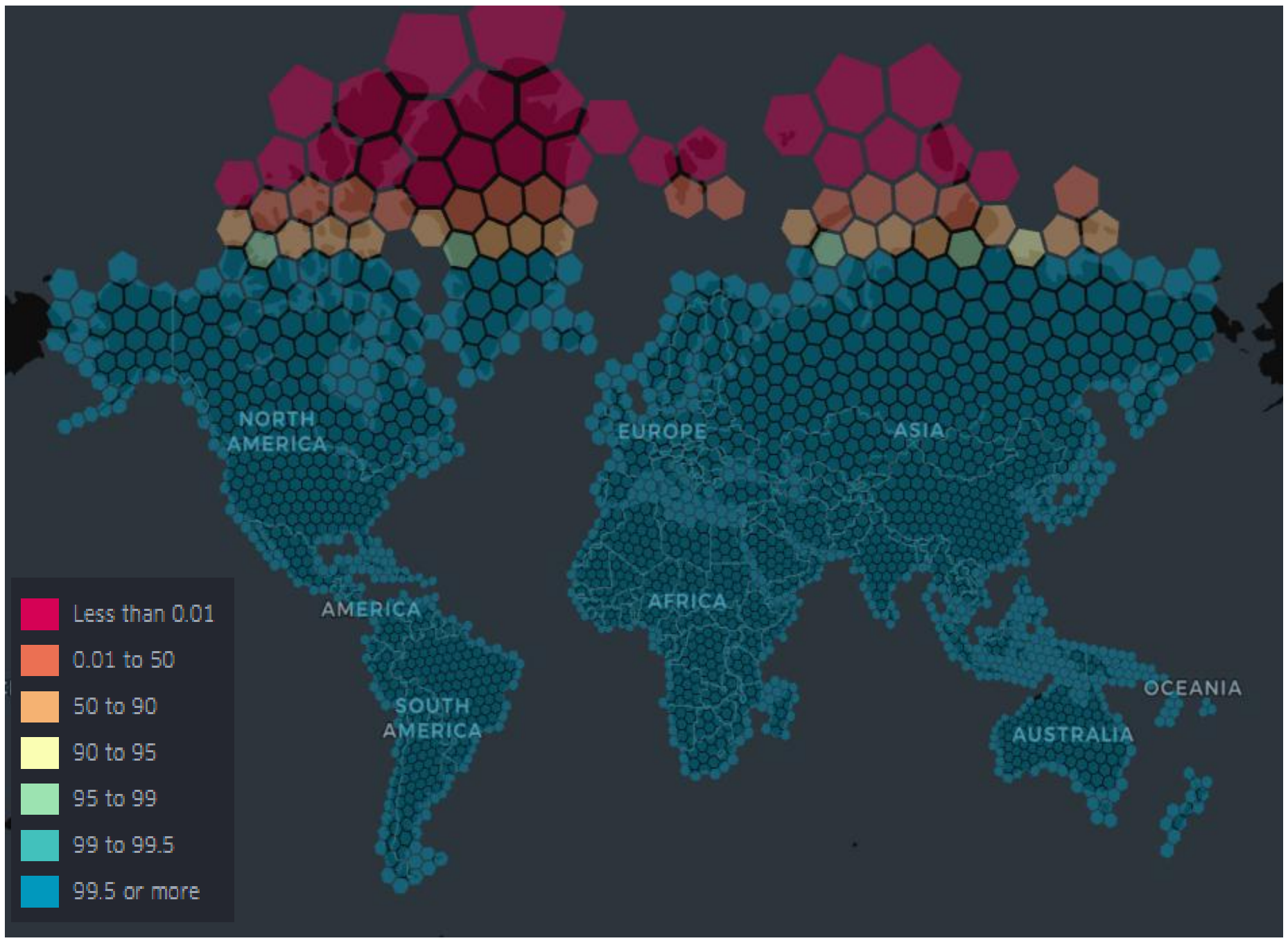

**FIGURE 5** Availability (%) of 4 Visible Pulsar Satellites Above 5° Elevation - 192 Satellites

At this stage of deployment, Pulsar can offer independent PNT services with HDOP close to or less than 1, as shown in Figure 6 enabling precise positioning performance. Figure 6 further illustrates the number of satellites in view as a function of latitude, reaching a maximum of 16 satellites at mid-latitudes. Both plots in Figure 6 show $50^{th}$, $68^{th}$, $95^{th}$, and $99^{th}$ percentiles to visualize the distribution of DOP and visibility throughout the duration of the simulation.

Comparing the two figures shows that the relationship between satellite count and geometry is non-linear: increasing the number of visible satellites from approximately 13 to 16 yields only a modest improvement in HDOP, whereas increasing the constellation from around 8 to more than 10 satellites results in a substantially larger reduction in HDOP. This demonstrates that the greatest gains in geometric strength are achieved when transitioning from lower to moderate satellite availability, with diminishing returns as the constellation becomes denser.

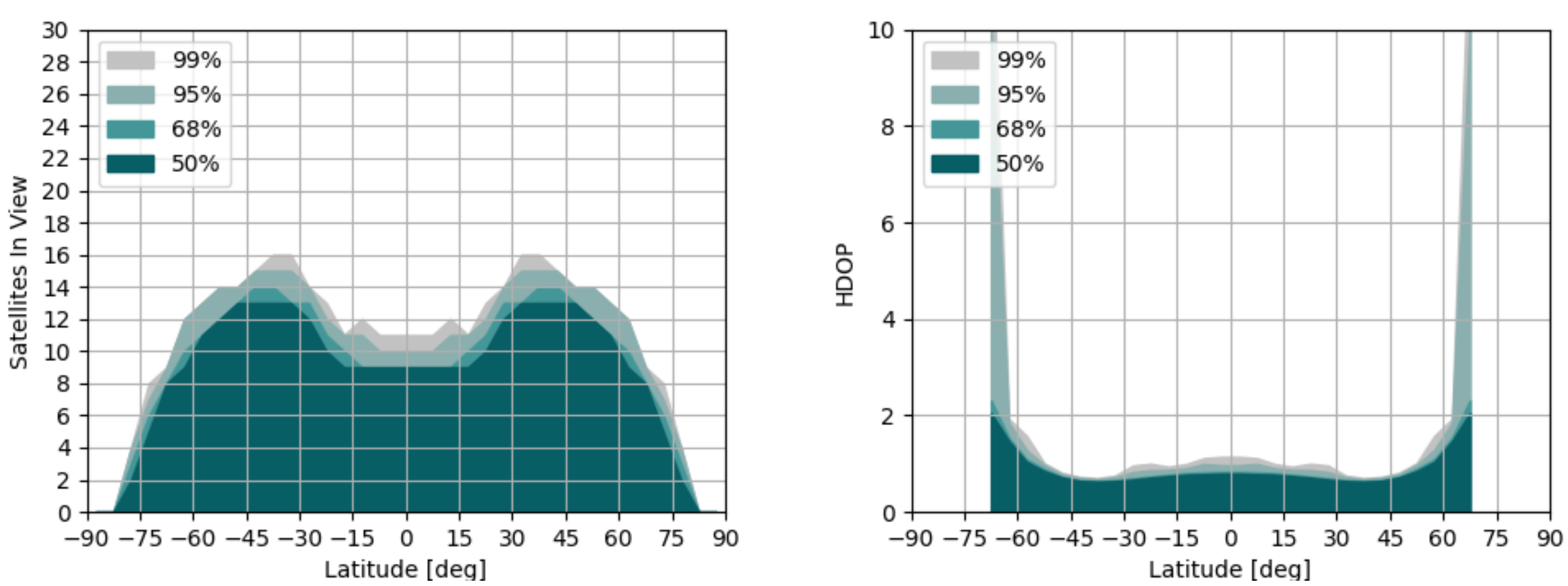

**FIGURE 6** Pulsar Satellite Visibility (left) and HDOP (right) by Latitude Above 5° Elevation - 192 Satellites

### 3.1.4 *Full Operational Capability: 258 Pulsar Satellites*

At full operational capacity, Pulsar consistently provides 4 satellites in view globally with added redundancy, even with the elevation mask raised to 10°, as shown in Figure 7.

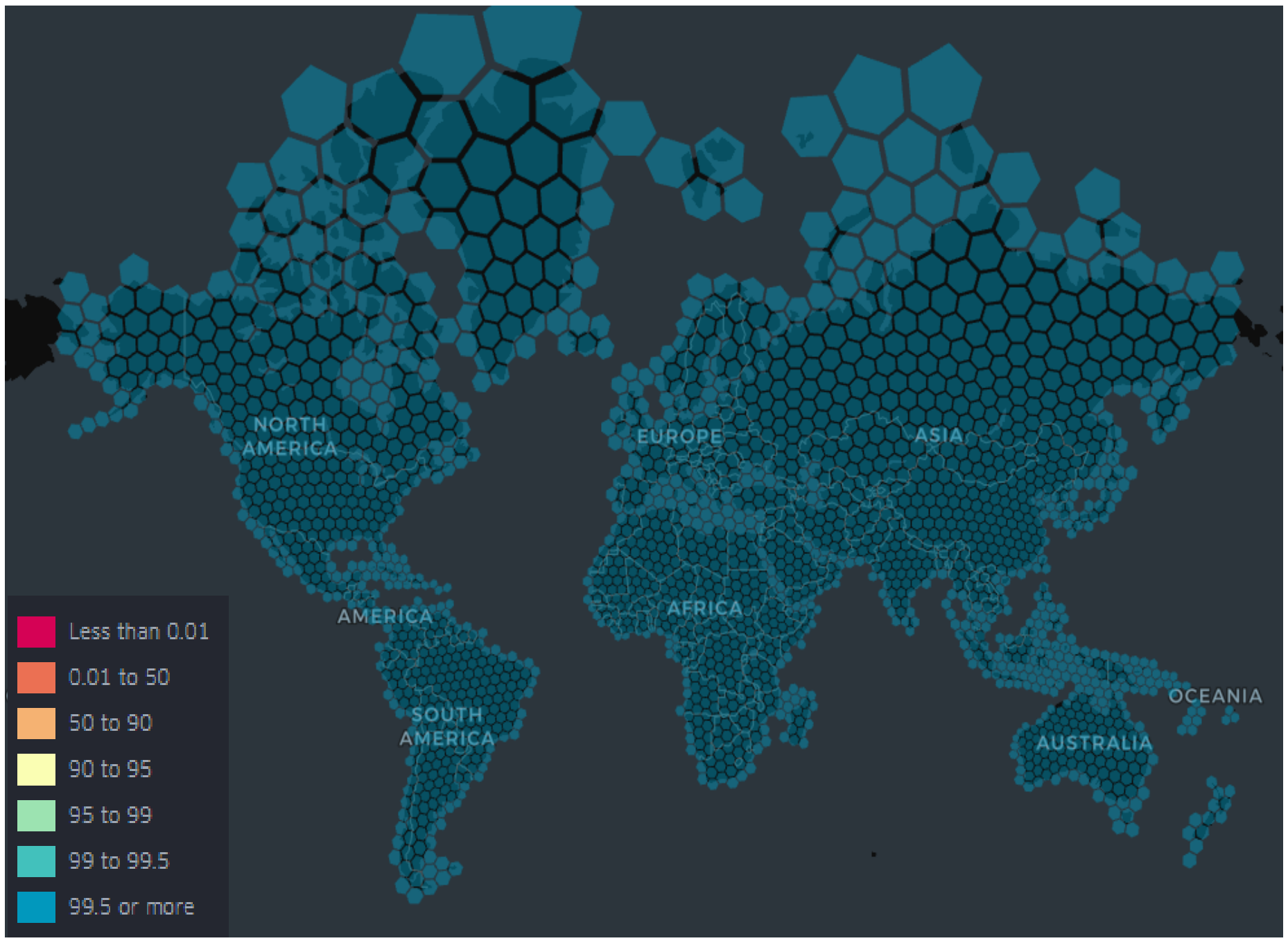

**FIGURE 7** Availability (%) of 4 Visible Pulsar Satellites Above 10° Elevation - 258 Satellites

Figure 8 shows the worst case for satellite visibility, defined as the minimum number of satellites visible for each simulation. Across all locations, the minimum number of satellites visible is 6, with some locations having minimums as high as 14 satellites. This demonstrates that Pulsar will be able to operate as a standalone PNT constellation, independent of other GNSS. The maximum number of visible satellites at any location is greater than 20.

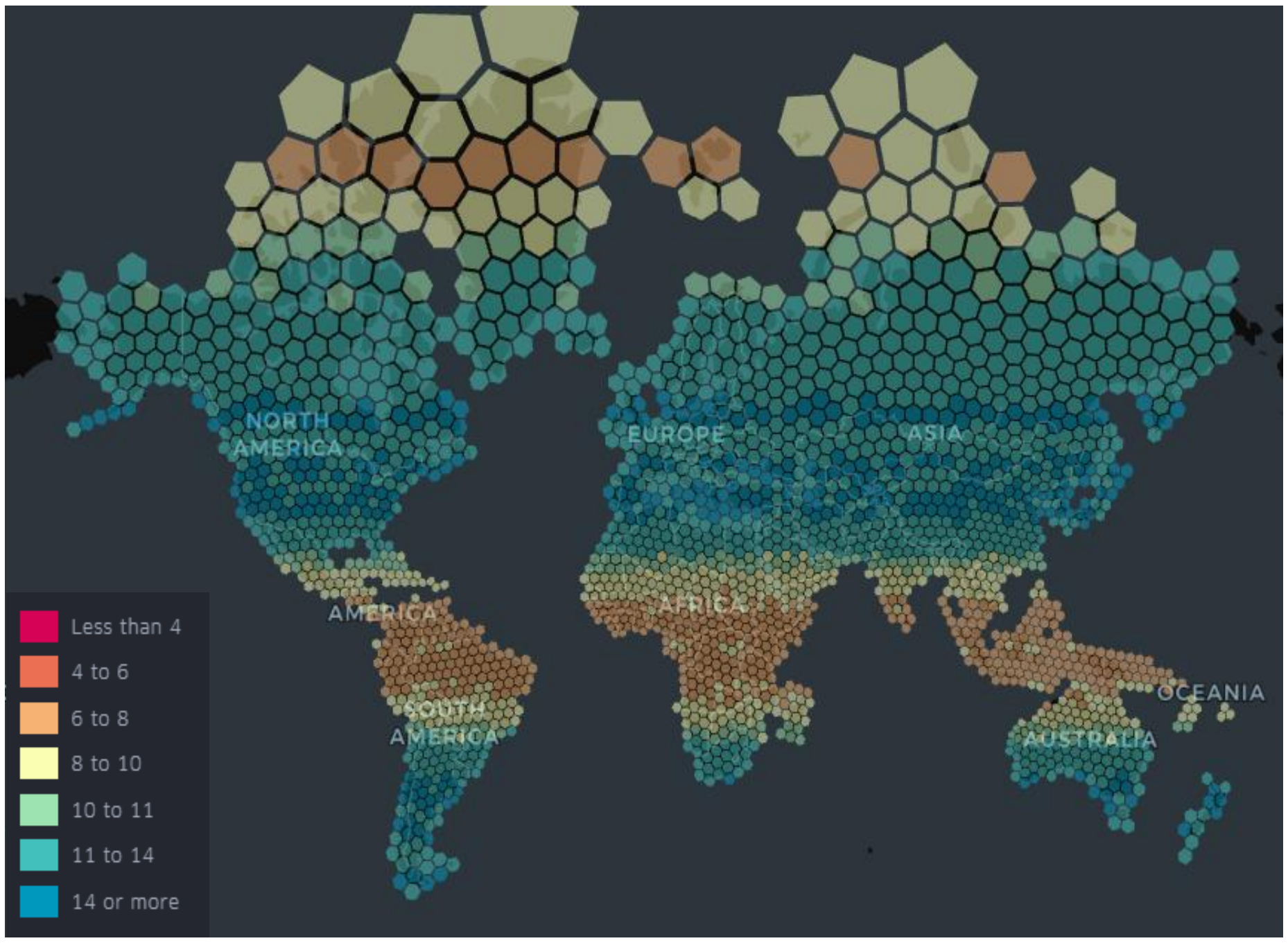

**FIGURE 8** Minimum Number of Visible Pulsar Satellites Above 5° Elevation at Full Operational Capability

The satellite visibility was broken down by latitude in Figure 9. On average, GPS will have 10 to 12 satellites visible at any given time, and limited availability at far north and far south latitudes. Figure 9 shows that Pulsar will have at least 10 satellites visible above 5° elevation, at all locations, 99% of the time.

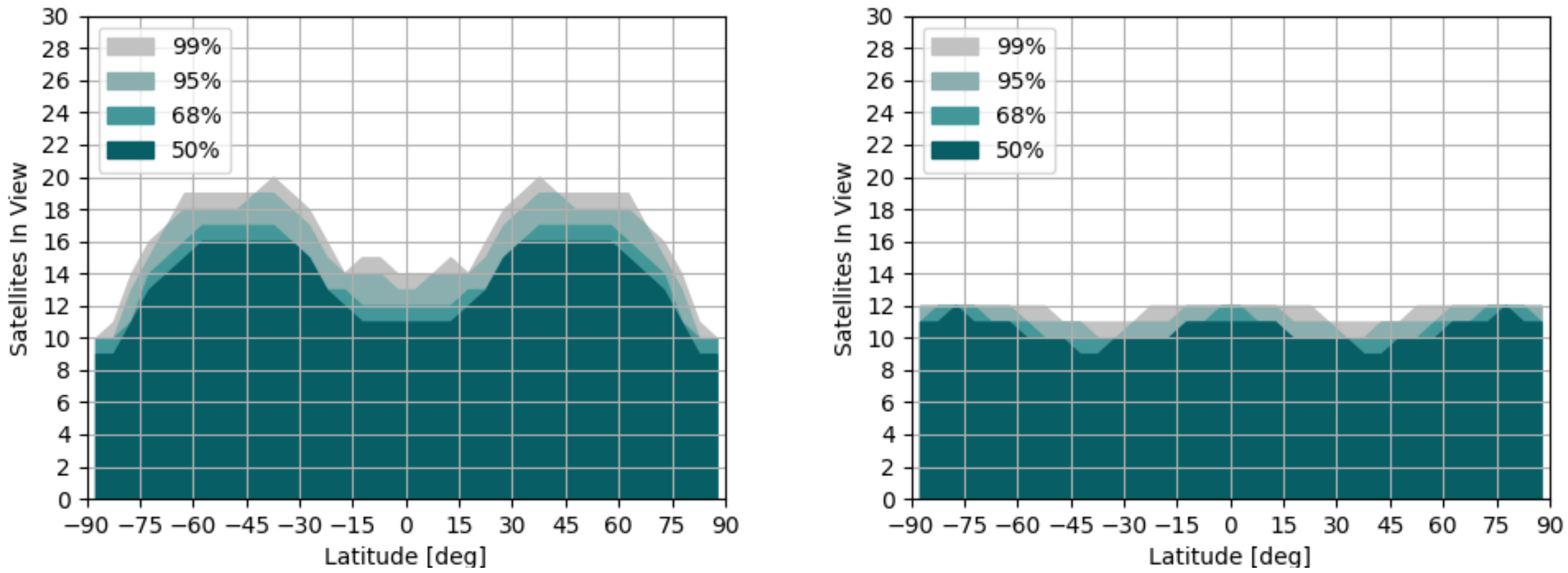

**FIGURE 9** Pulsar Satellite Visibility by Latitude Above 5° Elevation - 258 Pulsar Satellites (left) and GPS (right)

The HDOP and PDOP for Pulsar at full operational capability are shown in Figure 10. In practice, a PDOP of 6 or lower provides sufficient geometry for GPS position estimation. At all latitudes, Pulsar's PDOP falls comfortably below this threshold. These plots demonstrate that Pulsar should provide comparable DOP to GPS and therefore will have satellite geometry sufficient for single point positioning (SPP). In PPP mode, Pulsar will be even more advantageous due to the added measurement diversity and geometric variation.

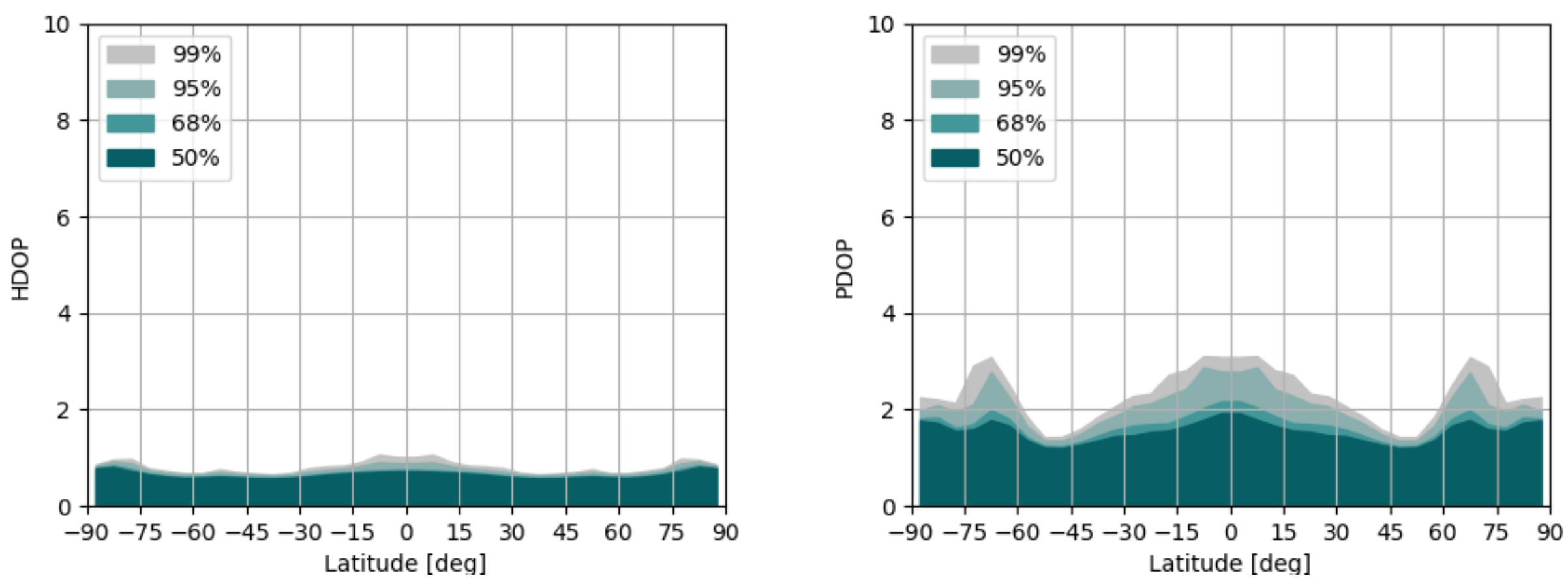

**FIGURE 10** Pulsar HDOP (left) and PDOP (right) by Latitude Above 5° Elevation - 258 Satellites

The increased satellite availability of Pulsar at each deployment stage is anticipated to unlock more PNT capabilities. At 48 satellites, most regions at low- and mid- latitudes will have at least one Pulsar satellite visible more than 99% of the time. At 192 satellites, at least four Pulsar satellites are visible more than 99.5% of the time at all locations, excluding the North and South poles. With the addition of the 66 polar satellites, Pulsar will have full global coverage with built-in redundancy.

### 3.2 PPP Convergence Simulations

This section will cover two cases of PPP convergence analyses, the first consisting of a global, software-based simulation, and the second consisting of a covariance analysis of data from Pulsar-0 processed through RTKLIB.

### *3.2.1 Part 1: PPP Convergence from Global Simulations*

The convergence performance of a PPP engine is dependent on several factors including the error correction methods, the quantity and quality of observations, and the estimation filter design. In this study, we implemented a simple version of a PPP estimator to evaluate the impact of Pulsar on positioning precision.

To the best of our knowledge, there are no commercially available positioning engines that have implemented PPP with LEO. This study is intended to demonstrate and predict the performance of PPP with Pulsar as an independent system and as a GNSS augmentation. The observations are generated within the simulation and therefore observation errors such as clock offsets and atmospheric delays can be resolved cleanly with minimal impact on position errors.

Figure 11 shows the 95$^{th}$ percentile convergence curves of horizontal error for over 4000 simulated PPP runs. The three greyscale lines are GPS, Galileo, and GPS + Galileo. The three teal-coloured lines are for Pulsar and MEO augmented with Pulsar. In this simulation, the full Pulsar constellation with 258 satellites is modelled alongside nominal GPS and Galileo constellations and pseudorange and carrier phase observations are generated and filtered through a simple PPP algorithm. The results in Figure 11 are from dual-frequency float solutions.

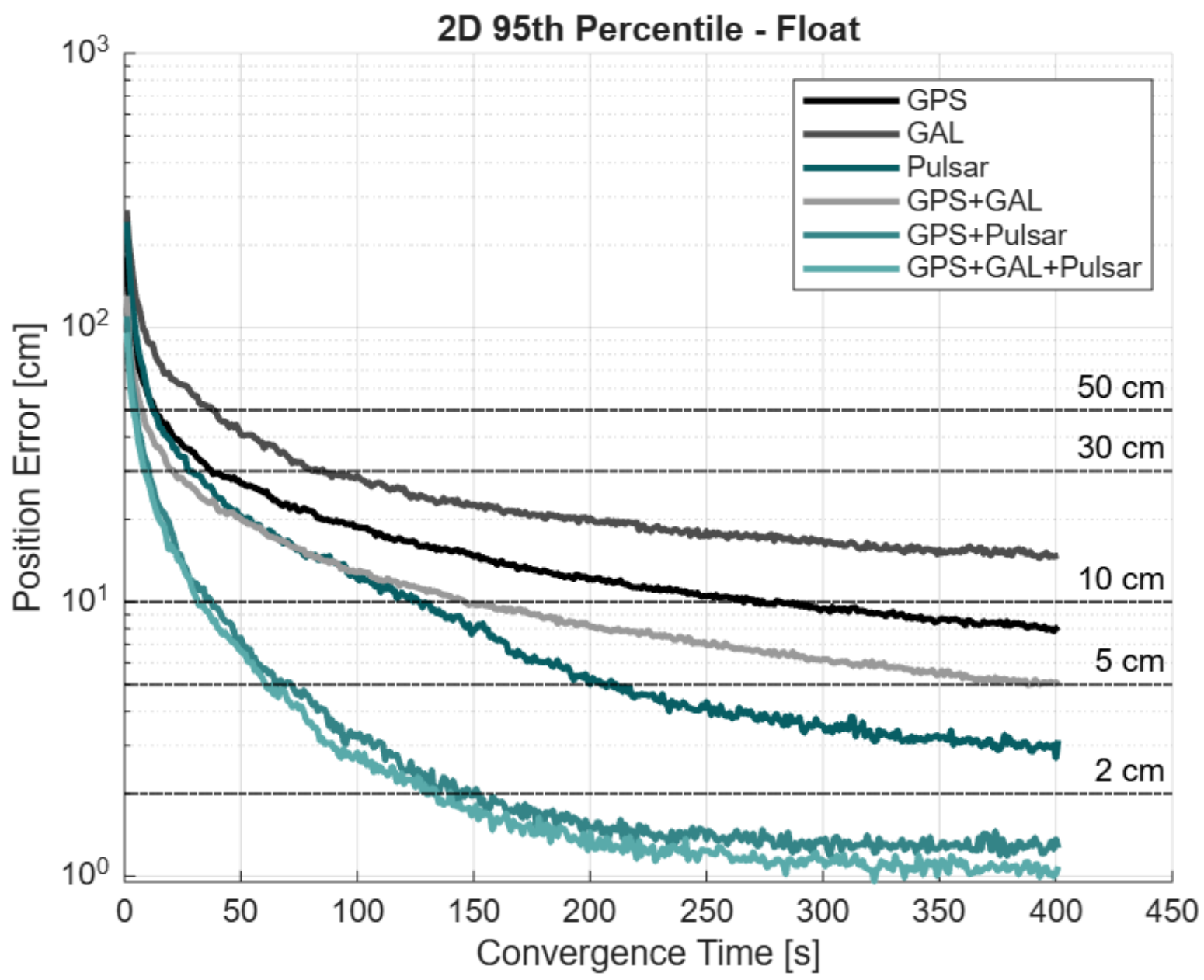

**FIGURE 11** 95th Percentile Horizontal Position Error Convergence from Numerical Simulations

A summary of 2D position error convergence times with dual frequencies and float ambiguities for GPS, Pulsar FOC, and GPS + Pulsar FOC to a threshold of 10 cm are provided in Table 3. Pulsar FOC reaches this threshold in less than half the time taken for GPS. With both GPS and Pulsar, 10 cm position error is achieved in only 38 seconds, 7× faster than GPS only.

**TABLE 3**

Summary of 2D 95$^{th}$ Percentile Position Error Convergence Times to 10 cm

| Constellations | 2D Error Convergence Time (s) |
|---|---|
| GPS | 266 |
| Pulsar FOC | 127 |
| GPS + Pulsar FOC | 38 |

Figure 12 shows the improvement in the time to 5 cm horizontal precision for each stage of the Pulsar constellation, compared to a baseline of dual-frequency GPS + Galileo. Each bar of this graph corresponds with the subsections in 3.1, where satellite availability is analyzed. As the satellite availability at an individual location is increased, more Pulsar satellites will be included in the PPP simulations, and the convergence time will improve. Additionally, as global coverage is expanded, improved PPP convergence will become available to more users.

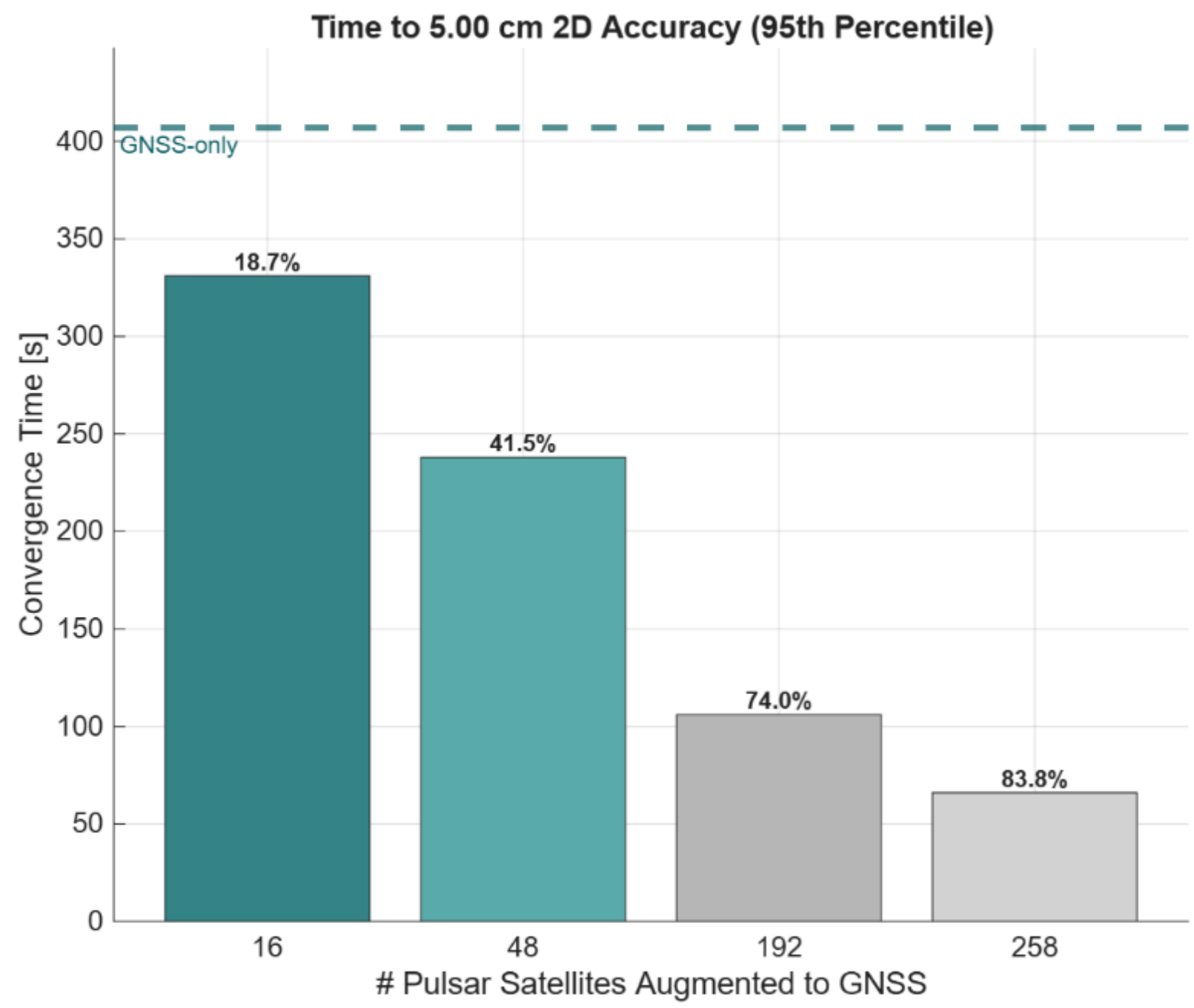

**FIGURE 12** Improvement in PPP Convergence Time with Pulsar Evolution from Numerical Simulations

The simulations in this section demonstrate the impact of Pulsar on the convergence of a PPP solution. At full operational capability, Pulsar improves the horizontal position error by 83%. Even with the MVP of 16 satellites, we still observe a meaningful improvement of 18% faster float solution convergence.

### *3.2.2 Part 2: PPP Convergence from Pulsar-0 with RTKLIB*

This section analyzes the horizontal and vertical standard deviations of a single frequency, GPS-only float solution augmented with Pulsar-0. In a similar manner to the earlier DOP analysis, the covariance matrix depends only on the design matrix in equation (1), and the previous state. This means the improvement in the covariance can be attributed to the faster movement of Pulsar-0 compared to GPS.

Figure 13 shows the convergence of the horizontal precision of GPS-only and GPS with Pulsar-0, Xona's IOV satellite over 2.5 minutes. Pulsar-0 has a lower orbital altitude and therefore faster motion than Pulsar's production constellation. This plot highlights the impact of the dynamic geometry of LEO satellites, showing significant improvement to the convergence curve with only 1 additional Pulsar satellite.

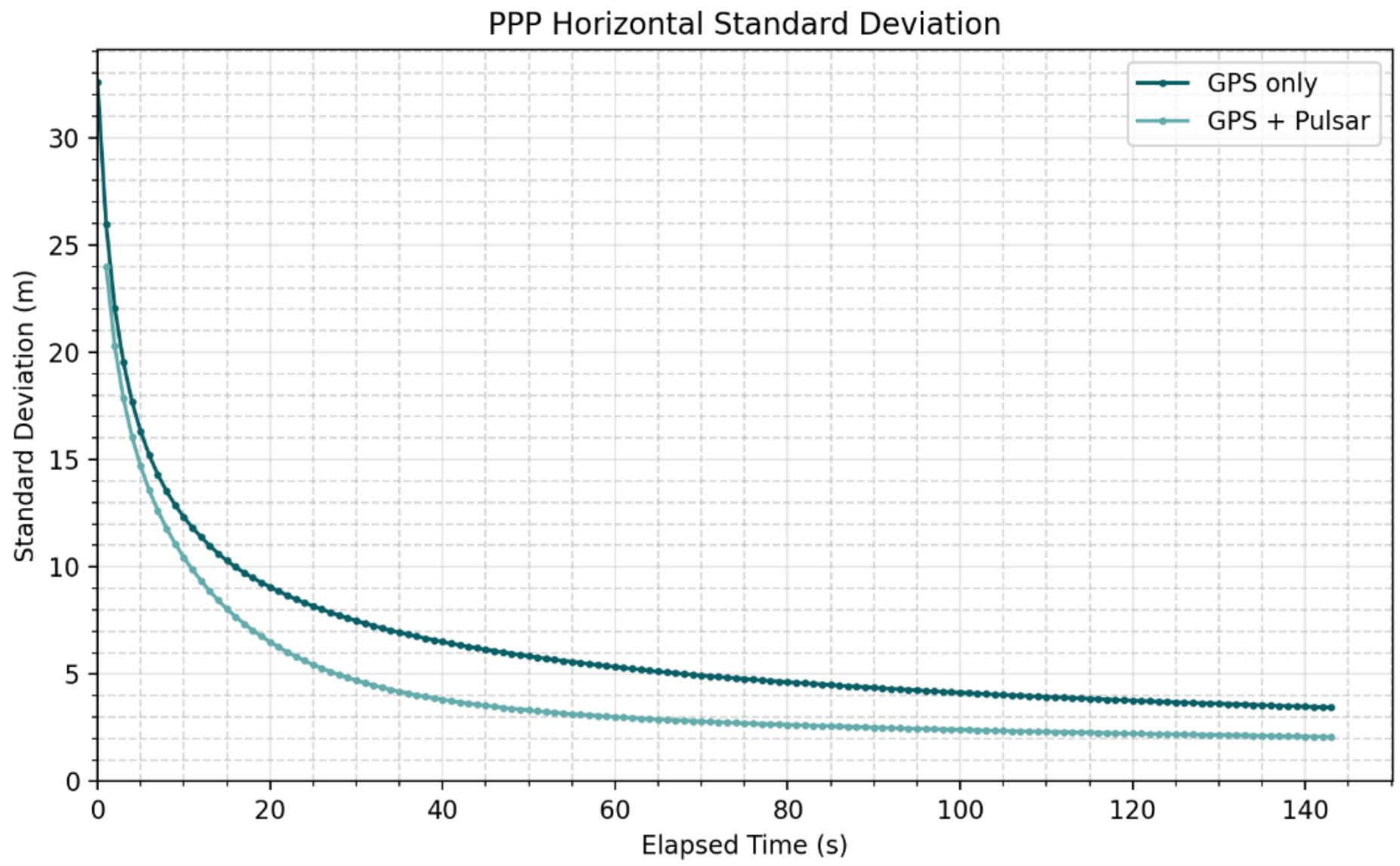

**FIGURE 13** Single Frequency, Horizontal Float Precision from IOV Satellite Positions and RTKLIB

In the horizontal plane, the float precision convergence is notably faster with the addition of Pulsar-0, with the standard deviation reaching 5 m in approximately half the time as it takes for GPS only. In the vertical direction shown in Figure 14, the initial variance is very high, but a similar improvement is evident as the GPS + Pulsar trendline drops further and faster than GPS only. After 2 minutes, the vertical standard deviation reaches 4.5 m for GPS only and 2.5 m for GPS+ Pulsar.

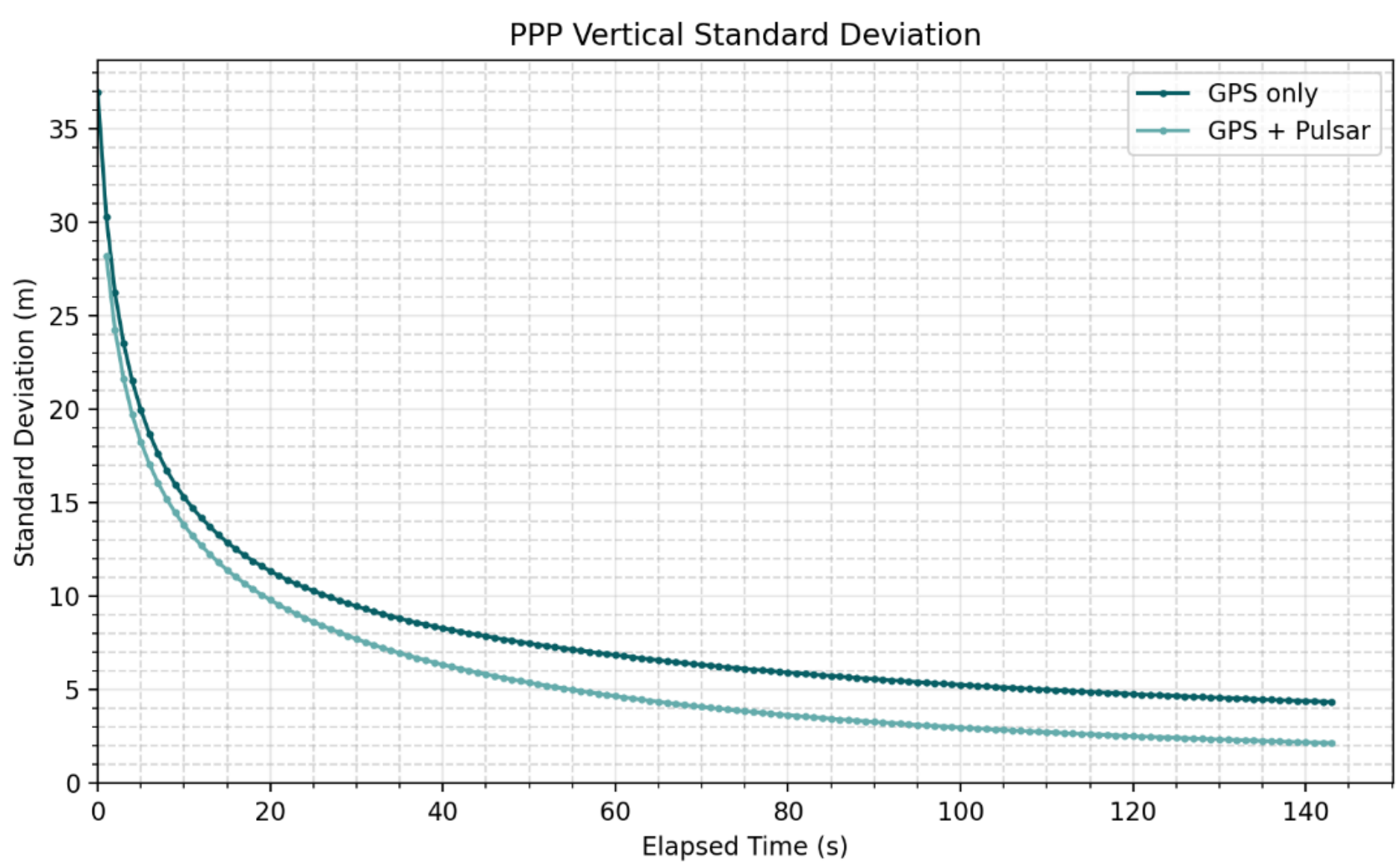

**FIGURE 14** Single Frequency, Vertical Float Precision from IOV Satellite Positions and RTKLIB

These results show faster convergence of the PPP filter in single-frequency float mode. Although ambiguity resolution has not been addressed here, better float precision translates to a higher likelihood that ambiguities can be resolved accurately. As stated earlier, PPP position error results based on real Pulsar-0 measurements are not presented because, with only one Pulsar satellite available, the PPP estimation problem is underdetermined. In particular, the receiver clock bias and the Pulsar system clock offset are not independently observable. However, the position precision or covariance is independent of observation errors and can therefore provide an idea of the impact of LEO satellites such as Pulsar-0 on PPP convergence under the assumption that appropriate corrections have been applied. The PPP filter in RTKLIB accepts and processes the observations from Pulsar-0 and reflects the added geometric strength through the covariance convergence.

## 4. CONCLUSIONS

In this paper, the availability and satellite geometry of Xona's LEO constellation, Pulsar, was analyzed at several stages of the constellation's deployment (16, 48,192, and 258 satellites) to evaluate the orbital design for positioning applications. With the inclined planes fully populated, horizontal and 3D position DOP was shown to be comparable to GPS at all locations except for far north and south latitudes. At full operational capability, Pulsar's 258 satellites in orbit will meet or exceed the availability of GPS globally.

The results of this paper support the expectation that the geometry of LEO GNSS will improve PPP convergence through simulations of Pulsar both independently and when combined with GPS. In Section 3.2.1, the PPP simulations show that the float solution convergence time is improved by combining GPS Pulsar at all 4 stages of deployment, with an 18.8% improvement with 16 Pulsar satellites and an 83.8% improvement with 258 Pulsar satellites In Section 3.2.2, data from Pulsar-0, processed through RTKLIB, showed significantly faster covariance convergence.

In interpreting the PPP convergence results, it is important to understand where error sources present themselves in results. This paper features two cases of PPP analysis. The first one is conducted entirely in simulation and assigns stochastic error to observations, leading to position error results that are reflective of the expected error magnitude and assume no major outliers or unmodelled biases. The second case of PPP analysis features real data from Pulsar's IOV satellite, which likely does contain errors and biases that are challenging to resolve, especially with only one satellite. However, in the second case, the satellite geometry can be separated from the observation errors by analyzing the covariance rather than the position error. While neither of these cases can claim to be fully representative of live PPP performance, the position error from simulated observations and the covariance from real satellite data together provide a compelling case for future PPP performance from Pulsar.

It will take some time for Pulsar to reach a stage where its satellite availability and positioning capabilities match existing GNSS constellations, but availability and DOP results show that the final constellation will surpass GPS in visibility, particularly in the mid-latitudes. Additionally, PPP convergence simulations have shown that as an augmentation to MEO GNSS, Pulsar can begin supporting better positioning performance as early as the 16-satellite stage. This paper has demonstrated that as the constellation grows, Pulsar will provide increasing value and performance improvements to PNT applications.

## 5. FUTURE WORK

Future work on this topic will undoubtedly feature live signal testing and positioning performance evaluations as more Pulsar satellites are launched. PPP simulations will also be expanded to include detailed statistical analysis identifying the best-case and worst-case satellite geometry for Pulsar, as well as how urban environments and high multipath conditions affect PPP convergence. Another closely related area of study will be generating corrections specifically for LEO satellites and applying them to PPP solutions, including a performance assessment of Pulsar's ranging errors.

From this new RTKLIB implementation, more research into positioning filters and algorithms will be possible including further testing involving carrier phase ambiguity resolution. Additionally, real-time positioning engines will have to be updated to accept new signals and constellations, as well as to tune their estimation filters to optimize LEO signals. Observation weighting and process noise, especially at low elevation angles, may need to be adapted to take full advantage of LEO constellations. There are many other receiver design considerations to be made in acquisition, tracking, and computational efficiency, as discussed in literature (Leclère et al., 2025) (Eissfeller et al., 2024). However, because Pulsar has been designed for backwards compatibility with existing GNSS receiver hardware, all these changes should be possible through firmware or software upgrades.